\documentclass{JAC2003}
 
\usepackage[dvipdfm]{graphicx}
\usepackage{amsmath}
\usepackage{amssymb}

\begin{document}
\title{Can we detect "Unruh radiation" in the high intensity lasers?
\thanks{The report  is based on a talk by S.Zhang and \cite{IYZ}.}}
\author{
Satoshi Iso\thanks{satoshi.iso@kek.jp}, 
Yasuhiro Yamamoto\thanks{yamayasu@post.kek.jp} and 
Sen Zhang\thanks{zhangsen@post.kek.jp}, KEK, Tsukuba, Japan}

\maketitle

\begin{abstract}
An accelerated particle sees the Minkowski vacuum as thermally excited,
which is called the Unruh effect.
Due to an interaction
with the thermal bath, the particle moves stochastically 
like the Brownian motion in a heat bath. 
It has been discussed that the accelerated charged particle may
emit extra  radiation (the Unruh radiation  \cite{ChenTajima}) 
besides the Larmor radiation, and
 experiments are under planning to detect such radiation 
by using ultrahigh intensity lasers \cite{ELI,ELIhp}.  
There are, however, counterarguments
that the radiation is canceled by an interference effect
between the vacuum fluctuation and the radiation from the 
fluctuating  motion.
In this and another reports \cite{IYZ2}, 
we review our recent analysis on the issue of the 
Unruh radiation. In this report, we particularly consider the 
thermalization of an accelerated particle in the scalar QED,
and derive  the relaxation time of the thermalization.
The interference effect is discussed separately
in  \cite{IYZ2}.

\end{abstract}

\section{UNRUH EFFECT AND UNRUH RADIATION}
Quantum field theories in the space-time with horizons exhibit 
interesting thermodynamic behavior.
The most prominent phenomenon is the Hawking radiation and 
the fundamental laws of thermodynamics hold in the black hole background.
A similar phenomenon  occurs for a uniformly accelerated observer 
in the ordinary Minkowski vacuum \cite{Unruh}.
This is called the Unruh effect.
If a particle is uniformly accelerated in the Minkowski space 
with an acceleration $a$, there is a causal horizon (the Rindler horizon) 
and no information can be transmitted from the other side of the horizon.
Because of the existence of the Rindler horizon, 
the accelerated observer sees the Minkowski vacuum as 
thermally excited with the Unruh temperature
\begin{align}
 T_U = \frac{\hbar a}{2 \pi c k_B} = 4 \times 10^{-23}
  \left( \frac{a}{1 \ \textrm{cm/s}^2} \right)[\textrm{K}].
\end{align}

Since the Unruh temperature is very small for ordinary acceleration, 
it was very difficult to detect the Unruh effect.
But with the ultra-high intensity lasers, 
the Unruh effect can be experimentally accessible.
In the electro-magnetic field of a laser with intensity $I$, 
the Unruh temperature is given by
\begin{align}
 T_U = 8 \times 10^{-11} \sqrt{\frac{I}{1\,\textrm{W/cm}^2}}\ [\textrm{K}].
\end{align}
The Extreme Light Infrastructure project \cite{ELIhp} is planning to 
construct Peta Watt lasers with an intensity 
as high as $5 \times 10^{26} \ [\textrm{W/cm}^2]$. 
The expected Unruh temperature becomes more than $10^3 \,\textrm{K}$.
So it is time to ask ourselves 
how we can experimentally observe such high Unruh temperature of 
an accelerated electron in the laser field.

\begin{figure}[htb]
\begin{center}
 \includegraphics[width=10em]{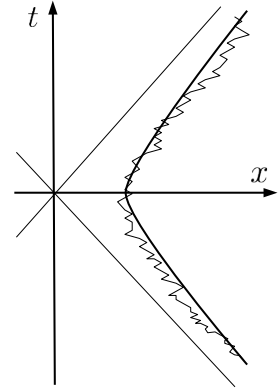}
\caption{Stochastic trajectories of a uniformly accelerated electron
affected by quantum field fluctuations.} 
 \label{Fluctuation}
\end{center}
\end{figure}

Chen and Tajima proposed that one may be able to detect 
the Unruh effect by observing quantum radiation \cite{ChenTajima}
from the electron.
It is called the Unruh radiation.
Since a uniformly accelerated electron feels the 
vacuum (with quantum virtual pair creations of particles and anti-particles)
as thermally excited with the Unruh temperature,
the motion of the electron fluctuates and is expected to
become thermalized(Fig.~\ref{Fluctuation}).
This fluctuating motion of the electron changes the acceleration 
of the electron and may produce  additional radiation (the Unruh radiation)
to the ordinary Larmor radiation.
The rough estimation \cite{ChenTajima} suggested that 
the strength of the Unruh radiation is much smaller than the classical one 
by $10^{-5}$, but the angular dependence becomes quite different.
Especially in the direction along the acceleration there is a blind spot for 
the Larmor radiation while the Unruh radiation is expected to be
radiated more spherically. 
Hence they proposed to detect the Unruh radiation by setting a photon detector
in this direction. 

The above argument seems intuitively correct, 
but there are two problems that should be clarified.
The first problem is the {\it thermalization time} of the fluctuation.
The electromagnetic field of laser are not constant but oscillating.
One may  approximate  the electron's motion around 
the turning points by a uniform acceleration.
This approximation is valid only when  the period of the laser 
is large enough compared to the relaxation time (or thermalization time)
of the particle's fluctuation.
Using a stochastic approach, we obtained the relaxation time of the fluctuation
and showed that the relaxation time is longer than the period of the laser.
In such a case, 
we must fully analyze the transient dynamics of the fluctuation 
to calculate the radiation in the laser field.

The second problem is  the {\it interference effect}.
Since the Unruh radiation originates in the interaction with the particle
with the quantum fluctuations of the vacuum, we cannot neglect
the interference  of the Unruh radiation and the vacuum quantum 
fluctuations. 
In a simpler  model, 
it has been known that the Unruh radiation is completely
canceled by the interference effect. 
The cancellation was shown for the Unruh detector in both 1+1 and 
3+1 dimensions\cite{Sciama,RavalHuAnglin}.
There was no calculation of the  interference effect in the 
case of the uniformly accelerated charged particle since the calculation
 needs some technicalities.
In the paper \cite{IYZ}
 we calculated the interference effect for the charged particle 
in the scalar QED and 
found that some of the Unruh radiation is  canceled by the interference
effect,  but the cancellation occurs only partially.
So we still have a possibility to detect additional radiation
from the uniformly accelerated charged particle,
but the complete understanding needs more detailed analysis.

In the rest of this report 
 we first briefly review the stochastic model of a uniformly accelerated
charged particle  and then show 
how the thermalization of the fluctuation occurs 
by solving the stochastic equation.
Finally we briefly sketch the calculation of the radiation, 
particularly put emphasis on the interference effect.
More details of the calculation of the interference effect and 
the Unruh radiation are 
reviewed in another report of the same authors in the proceedings
\cite{IYZ2}.

\section{THERMALIZATION}
We consider the scalar QED. 
The model is analyzed in~\cite{JohnsonHu} and 
here we briefly review the settings and the derivation of 
the stochastic Abraham-Lorentz-Dirac (ALD) equation.
The system composes of a relativistic particle $z^\mu(\tau)$ and the scalar field $\phi(x)$. The action is given by
\begin{align}
 S[z,\phi, h] =& 
  -m \int d\tau \, \sqrt{\dot{z}^{\mu} \dot{z}_\mu}
  +\int d^4x \, \frac{1}{2}(\partial_\mu \phi)^2 \nonumber \\
& + \int d^4x \, j(x;z) \phi(x) .
\end{align}
The scalar current $j(x;z)$ is defined as
\begin{align}
 j(x;z) = e \int d\tau \, \sqrt{\dot{z}^\mu \dot{z}_\mu}\, \delta^4 (x-z(\tau)),
 \label{current}
\end{align}
We choose the parametrization $\tau$ to satisfy $\dot{z}^2=1$.

\subsection{The Stochastic Equation}
The equation of motion of the particle is given by
\begin{align}
 m\ddot{z}^\mu = 
   F^\mu - \int d^4 x \frac{\delta j(x;z)}{\delta z_\mu (\tau)} \phi(x)
\label{particle-eom}
\end{align}
where we have added the external force $F^\mu$ so as to accelerate the particle uniformly;
$F^\mu = ma(\dot{z}^1, \dot{z}^0, 0,0)$.
Then a classical solution of the particle (in the absence of
 the coupling to $\phi$) is given by
\begin{align}
  z^\mu_0 = (\frac{1}{a}\sinh{a\tau},\frac{1}{a}\cosh{a \tau},0,0).
 \label{classical-trajectory}
\end{align}
The equation of motion of the radiation field 
$ \partial^2 \phi = j$
is  solved by using the retarded Green function 
$G_R$ as
\begin{align}
 \phi(x) &= \phi_{h} (x) + \phi_{I}(x), \nonumber \\
 \phi_{I}(x) &= \int d^4 x' G_R(x,x') j(x';z) 
 \label{sol-phi}
\end{align}
where $\phi_h$ is the homogeneous solution of the equation of motion and 
represents the vacuum fluctuation.
It is responsible
for the particle's fluctuating motion under a uniform acceleration.
Inserting the solution (\ref{sol-phi}) into (\ref{particle-eom}),
we have the following stochastic equation for the particle
\begin{align}
 m\ddot{z}^\mu(\tau) =& 
   F^\mu(z(\tau)) -e \vec{\omega}^\mu \label{stochastic} \\
&\times \left( 
   \phi_h(z(\tau)) + e \int d\tau' \ G_R(z(\tau), z(\tau'))  
  \right),
\end{align} 
where 
$\vec{\omega}_\mu = \dot{z}^\nu \dot{z}_{[\nu} \partial_{\mu]} - \ddot{z}_\mu$,
which comes from the deviation of the current 
\begin{align}
 \int d^4 x \frac{\delta j(x;z)}{\delta z^\mu (\tau)} f(x) = 
  e\vec{\omega}_\mu f(x)|_{x=z(\tau)}.
\end{align}

The homogeneous part $\phi_h(z(\tau))$ of the scalar field describes 
the Gaussian fluctuation of the vacuum, hence, 
the first term in the parenthesis of (\ref{stochastic}) can be 
interpreted as random noise to the particle's motion
\begin{align}
 \langle \phi_h(x) \phi_h (x') \rangle 
 = -\frac{1}{4 \pi^2}  \frac{1}{(t-t'-i\epsilon)^2-r^2}.
\label{phi-correlation} 
\end{align} 
It is essentially quantum mechanical, 
but if it is evaluated on a world line of a uniformly accelerated particle 
$x=z(\tau), x'=z(\tau')$, 
it behaves as the ordinary finite temperature noise. 
The second term in the parenthesis of (\ref{stochastic}) is 
a functional of the total history of the particle's motion 
$z(\tau')$ for $\tau' \le \tau$, but it can be reduced to 
the so called radiation damping term of 
a charged particle coupled with radiation field.
It is generally nonlocal, 
but since the Green function damps rapidly as a function of the distance $r$, 
the term is approximated by local derivative terms. 
After the mass renormalization, 
we get the following generalized Langevin equation for the charged particle,
\begin{align}
 m \dot{v}^\mu - F^\mu - \frac{e^2}{12\pi} (v^\mu \dot{v}^2 + \ddot{v}^\mu) =
  -e \vec{\omega}^\mu \phi_h(z)
\label{stochastic-particle}
\end{align}
where $v^\mu = \dot{z}^\mu$.
This equation is an analog of the ALD equation for a charged particle 
interacting with the electromagnetic field.
The dissipation term is induced by the effect of the backreaction of 
the particle's radiation to the particle's motion. 
Note that, if the noise term is absent, the classical solution 
(\ref{classical-trajectory}) with a constant acceleration is 
still a solution to the equation (\ref{stochastic-particle}).

\subsection{Equipartition Theorem}
The stochastic equation (\ref{stochastic-particle}) is nonlinear and difficult to solve.
Here we consider  small fluctuations around the classical trajectory induced by the vacuum fluctuation $\phi_h$. 
Especially we consider fluctuations in the transverse directions.
First we expand the particle's motion around the classical trajectory 
$z^\mu_0$ as
\begin{align}
 z^\mu(\tau) = z^\mu_0 (\tau) + \delta z^\mu (\tau).
\end{align} 
The particle is accelerated  along the $x$ direction.
In the following we consider small fluctuation in transverse directions.
By expanding the stochastic equation (\ref{stochastic-particle}), 
we can obtain a linearized stochastic equation for
the transverse velocity fluctuation $\delta v^i \equiv \delta \dot{z}^i$ as,
\begin{align}
 m \delta \dot{v}^i = 
 e \partial_i \phi_h + \frac{e^2}{12\pi} 
(\delta \ddot{v}^i -a^2 \delta v^i) .
\label{stochastic-transverse}
\end{align}
Performing the  Fourier transformation with respect to the trajectory's 
parameter
$\tau$
\begin{align}
 \delta v^i(\tau) &= \int \frac{d \omega}{2\pi} 
 \delta \tilde{v}^i(\omega) e^{-i \omega \tau}, \\ 
 \partial_i \phi_h(\tau) &= 
   \int \frac{d \omega}{2\pi} \partial_i \varphi(\omega) 
	     e^{-i \omega \tau},
\label{varphi}
\end{align}
the stochastic equation can be solved as
\begin{align}
 \delta \tilde{v}^i(\omega) = e h(\omega) \partial_i \varphi(\omega),
\end{align}
 where 
\begin{align}
 h(\omega) = \frac{1}{-im \omega +\frac{e^2(\omega^2 +a^2)}{12\pi}}.
\end{align}

The vacuum 2-point function along the classical 
trajectory can be evaluated  from (\ref{phi-correlation}) as
\begin{align}
 & \langle \partial_i \phi_h(x) \partial_j \phi_h (x') \rangle 
   |_{x=z(\tau),x'=z(\tau')} \nonumber \\
 &= \frac{1}{2\pi^2} 
     \frac{\delta_{ij}}{((t-t'-i \epsilon)^2 -r^2)^2} \nonumber \\
 &= \frac{a^4}{32 \pi^2}  
     \frac{\delta_{ij}}{\sinh^4(\frac{a(\tau-\tau'-i \epsilon)}{2})}.
\label{phi-correlation2} 
\end{align}
It has originated from the quantum fluctuations of the vacuum, 
but it can be interpreted as finite temperature noise 
if it is evaluated on the accelerated particle's trajectory \cite{Unruh}.
The Fourier transformation of the symmetrized two point function is evaluated as
\begin{align}
\langle \partial_i \phi(x) \partial_j \phi(x')\rangle_S
&= \langle \{\partial_i \phi(x), \partial_j \phi(x') \} \rangle /2 \nonumber \\ 
&= 2 \pi \delta(\omega+\omega') \delta_{ij} I_S (\omega),
\end{align}
where
\begin{align}
 I_S (\omega)= 
 \frac{1}{12 \pi} \coth \left(\frac{\pi \omega}{a}\right)
  (\omega^3 + \omega a^2),
\end{align}
which is an even function of $\omega$.
The correlator $I_S(\omega)$ should be regularized at the UV,
which is large $\omega$ or short proper time difference, 
where quantum field theoretic effects of electron become important.
Full QED treatment is necessary there.

For small $\omega$, it is expanded as
\begin{align}
I_S(\omega)=\frac{a}{12\pi^2} (a^2 + O(\omega^2)).
\end{align}
The expansion corresponds to the derivative expansion 
\begin{align}
 \langle \partial_i \phi_h(x) \partial_j \phi_h (x') \rangle_S 
 = \frac{a^3}{12\pi^2} \delta_{ij} \delta (\tau-\tau') 
  + \cdots .
\end{align}
With this expansion, the expectation value of the square of 
the transverse velocity fluctuation can be evaluated as
\begin{align}
 & \langle
    \delta v^i(\tau) \delta v^j(\tau^\prime)
  \rangle_S \nonumber \\
  &= 
  e^2 \int \frac{d \omega d\omega'}{(2\pi)^2} 
  \langle
    \partial_i \varphi(\omega) \partial_j \varphi(\omega')
  \rangle_S \ 
  h (\omega) h (\omega') e^{-i (\omega \tau+\omega' \tau')}
  \nonumber \\
  &\sim 
  e^2 \delta_{ij} \int \frac{d \omega }{24 \pi^3} 
  \frac{a^3 e^{-i \omega (\tau -\tau')}}
       {(m \omega)^2+\left(\frac{e^2}{12\pi}\right)(\omega^2+a^2)^2} .
\label{momentum-fluctuation}
\end{align} 
Here we consider the acceleration of the electron to be at the order 0.1 eV, 
which is much smaller than the electron mass 0.5 MeV.
With the assumption $m \gg a$, 
one can evaluate the integral and get the following result,
\begin{align}
\frac{m}{2}
\langle
  \delta v^i(\tau) \delta v^j(\tau)
\rangle
= \frac{1}{2}\frac{a \hbar}{2\pi c} \delta_{ij}
  \left( 1 + O (a^2/m^2) \right).
\label{equipartition}
\end{align} 
Here we have recovered $c$ and $\hbar$.
This gives the equipartition relation for the transverse 
momentum fluctuations in the Unruh temperature $T_U=a \hbar /2\pi c$. 

\subsection{Relaxation Time}
The thermalization process of the
stochastic equation (\ref{stochastic-transverse}) can be also 
discussed. For simplicity,  we  approximate the stochastic
equation by dropping the 
second derivative term. Then it is solved as
\begin{align}
 \delta v^i(\tau) =& e^{-\Omega_- \tau}\delta v^i(0) \notag \\ &
 +\frac{e}{m} \int_0^\tau d\tau^\prime \ \partial_i \phi(z(\tau')) 
   e^{-\Omega_- (\tau-\tau^\prime)},
\end{align}
where $\Omega_-$ is given by
\begin{align}
 \Omega_- = \frac{a^2 e^2}{12\pi m}
\end{align}
The relaxation time is $\tau_R=1/\Omega_-.$
The velocity square can be also calculated as
\begin{align}
 \langle
  \delta v^i(\tau) \delta v^j(\tau) 
 \rangle
 =& e^{-2 \Omega_- \tau} \delta v^i(0) \delta v^j(0) \notag \\ &
 + \frac{a \delta_{ij}}{2\pi m} (1- e^{-2 \Omega_- \tau}).
\end{align}
For $\tau \rightarrow \infty$, it approaches the thermalized average
(\ref{equipartition}). 
The relaxation time in the proper time
can be estimated, for the parameter $a = 0.1$ eV and $m=0.5$ MeV,
\begin{align}
 \tau_R  = \frac{12\pi m}{a^2 e^2} \sim 10^{-5} \textrm{sec}.
\end{align}
Let's compare this relaxation time with the laser frequency. 
The planned wavelength of the laser at ELI is around $10^3 $nm and 
the oscillation period of the laser field is very short; 
$3 \times 10^{-15}$ seconds.
So the relaxation time is much longer and 
the charged particle cannot become thermalized during each oscillation. 
Hence the assumption of the uniform acceleration in the laser field is not good.
Even in such a situation, if the electron is accelerated in the laser field 
for a long time, an electron may feel an averaged temperature. 

The position of the particle in the transverse directions 
also fluctuates like the ordinary 
Brownian motion in a heat bath.
The mean square of the transverse coordinate 
$R^2(\tau) =
 \sum_{i=y,z} \langle (z^i(\tau)-z^i(0))^2 \rangle $
is calculated as
\begin{align}
 R^2(\tau) &=
 2D \left( 
    \tau - \frac{3-4 e^{-\Omega_- \tau} + e^{-2\Omega_- \tau}}{2\Omega_-} 
  \right).
\end{align}
with the diffusion constant  
$ D = 2T_U/(\Omega_- m) = 12/a e^2 \sim 8 \times 10^4 \ m^2/s .$
In the Ballistic region where $\tau < \tau_R$, the mean
square becomes 
$ R^2(\tau) = 2T_U \tau^2 /m$
 while in the diffusive region 
($\tau > \tau_R$), it is proportional to the proper time
as 
$R^2(\tau) = 2D\tau. $
As the ordinary Brownian motion, the mean square of the particle's
transverse position grows linearly with time.
If it becomes possible to accelerate the particle 
for a sufficiently long period, it may be possible 
to detect such a Brownian motion in future laser experiments.

\section{RADIATION and INTERFERENCE}
Now we are ready to calculate the radiation emanated from 
the uniformly accelerated charged particle.
An important point is 
the interference effect 
between the quantum fluctuations of the vacuum $\phi_h$
and the radiation induced by the fluctuating motion in the transverse
directions $\phi_I$. 
First let's consider the two point function
\begin{align}
& \langle \phi(x) \phi(x') \rangle - \langle \phi_h(x) \phi_h(x') \rangle \\
&= \langle \phi_{I}(x) \phi_h(x') \rangle
+ \langle \phi_h(x) \phi_{I}(x') \rangle 
+\langle \phi_{I}(x) \phi_{I}(x') \rangle
\nonumber .
\end{align}
The Unruh radiation estimated in \cite{ChenTajima} corresponds to 
calculating the 2-point correlation function of the inhomogeneous terms 
$\langle \phi_{I} \phi_{I} \rangle$. 
(The same term also contains the Larmor radiation.)
However, this is not the end of the story.
As it has been discussed  in \cite{Sciama},
the interference terms  
$\langle \phi_{I} \phi_h \rangle + \langle \phi_h \phi_{I} \rangle$
may possibly  cancel the Unruh radiation 
in $\langle \phi_{I} \phi_{I} \rangle$ after the thermalization 
occurs. 
This is shown for an internal detector, 
but it is not obvious whether the same cancellation occurs 
for the case of a charged particle we are considering.

The energy-momentum tensor of the radiation field can be obtained 
from the 2-point function
\begin{align}
  \langle T_{\mu\nu} \rangle 
  = 
 \langle :\partial_{\mu}\phi \partial_{\nu}\phi 
 -\frac{1}{2} g_{\mu\nu} \partial^{\alpha}\phi \partial_{\alpha}\phi: \rangle_S.
\end{align}
It is written as a sum of the classical part 
 and the fluctuating part $T_{\mu\nu}=T_{cl,\mu \nu}+T_{fluc,\mu \nu}$.
The classical part corresponds to the Larmor radiation
while the fluctuating part contains both of the Unruh radiation and 
the interference terms. 

In \cite{IYZ} we calculated the 2-point function including 
the interference term,
and obtained the energy-momentum tensor.
The result we have obtained is summarized in \cite{IYZ2} in this proceedings.
Some terms are partially canceled  but not all. 
Hence, it seems that the uniformly accelerated charged particle
emits additional radiation besides the Larmor radiation.
The remaining terms after the partial cancellation 
are proportional to $a^3$ and 
suppressed  compared to the Larmor
radiation.
It has a different angular distribution,
but the additional radiation also vanishes in the forward direction.

\section{SUMMARY}
We have systematically
studied the thermalization of a uniformly accelerated charged particle
in the scalar QED using the stochastic method, and calculated the radiation by the particle. Two main messages of are
\begin{enumerate}
 \item "Long relaxation time compared to the laser period" 
 \item "Importance of the interference"
\end{enumerate}
The fluctuation of the particle doesn't become thermalized 
during the period of the laser,
and we need to study transient dynamics to obtain the radiation from 
an electron accelerated in the oscillating laser field.
The issue of the interference terms are more subtle. 
Since the particle fluctuation originates in the quantum vacuum fluctuation 
of the radiation field, it can be  by no means neglected.
Our result shows that the interference term 
partially cancels the Unruh radiation,
but some of them survives. The remaining Unruh radiation is smaller 
compared to the Larmor radiation
by a factor $a$ (acceleration) and has a different angular distribution. 
In this sense, it is qualitatively consistent with the proposal 
\cite{ChenTajima}.
But as we briefly review in \cite{IYZ2}, 
the additional radiation also vanishes in the forward direction.
and it seems difficult to detect such additional radiation experimentally
so far as the transverse fluctuation is concerned.  
Please beware  that the longitudinal fluctuations which we have not calculated yet
(because of technical difficulties related to a choice of gauge) 
may change the situation.


\end{document}